\documentclass{aa}
\usepackage[varg]{txfonts}
\usepackage{graphicx}
\usepackage{longtable}%
 \usepackage{lscape}
\usepackage{multicol} %
\usepackage{multirow}%
\usepackage{booktabs}%

\begin{document}

   \title{Gamma-ray and optical oscillations of  0716+714,
   MRK\,421, and BL\,Lac}


    \author{A. Sandrinelli \inst{1,2}, 
                S. Covino \inst{2},
                A. Treves \inst{1,2,3},
                E. Lindfors \inst{4},
                C. M. Raiteri \inst{5},
                K. Nilsson \inst{6}, 
                L. O. Takalo \inst{4}, 
                R. Reinthal \inst{4},
                A. Berdyugin \inst{4},
                V. Fallah Ramazani \inst{4},
                V. Kadenius \inst{4}, 
                T. Tuominen \inst{4}, 
                P. Kehusmaa \inst{7},  
                R. Bachev \inst{8},
                A. Strigachev\inst{8}           
                }

   \institute{Universit\`a dell'Insubria, Dipartimento di Scienza ed Alta Tecnologia,
              Via Valleggio 11, I-22100, Como, Italy      
             \and
             INAF-Istituto Nazionale di Astrofisica, Osservatorio Astronomico di Brera, 
             Via Bianchi 46, I-23807 Merate (LC), Italy      
             \and 
             INFN - Istituto Nazionale di Fisica Nucleare, Sezione Trieste - Udine,  Via Valerio 2, I-34127 Trieste, Italy         
            \and 
             Tuorla Observatory, Department of Physics and Astronomy, University of Turku, Finland         
            \and 
             INAF -  Istituto Nazionale di Astrofisica, Osservatorio Astrofisico di Torino, via Osservatorio 20, I-10025 Pino Torinese, Italy       
            \and
             FINCA - Centre for Astronomy with ESO, University of Turku, Finland
             \and
             Harlingten New Mexico Observatory, USA      
             \and
             Institute of Astronomy, Bulgarian Academy of Sciences, 72 Tsarigradsko shosse Blvd., 1784 Sofia, Bulgaria             
             \\
             \email{asandrinelli@yahoo.it}
              }

   \date{Received ...; accepted ....}

 
\abstract
{We examine the 2008-2016 $\gamma$-ray and optical light curves of three
 bright  BL Lac objects, 0716+714,   
 MRK\,421, BL\,Lac, which
 exhibit large structured  variability.  
We searched for periodicities by using a fully Bayesian approach.
 For two out of three sources investigated, no significant periodic variability was  found.
In the case of BL Lac,  we detected a periodicity of  $\sim$ 680 days.
Although the signal related to this is modest, the coincidence of the periods 
in both gamma   and optical bands  is indicative of a physical relevance.
Taking into consideration previous literature results, possibly related  $\gamma$-ray 
and optical periodicities of about one year time scale are proposed in 
four bright $\gamma$-ray 
blazars out of the ten  examined in detail. 
Compared with results from  periodicity search of optical archives of 
quasars, the presence of quasi-periodicities in blazars may be more 
frequent by a large factor. 
This suggests the intriguing possibility that the basic conditions 
for their observability are related to the relativistic jet in the observer direction, 
but the overall picture remains uncertain.}

   {}


 \titlerunning{Oscillations in Three  Blazars in the Northern Sky}
\authorrunning{A. Sandrinelli, S. Covino, A. Treves} 
  \maketitle
%

\section{Introduction}

\label{sec:intr}

  \begin{table*}
\begin{center}
\centering
\caption{ BL Lac objects considered in this study. 
}
\label{targets}
\small
\begin{tabular}{lcccc}
\hline\hline
Name            &ra                     &dec                    &z                      &R                 \\      
                        &[h m s]                &[d m s]                &                       &[mag]  \\
\hline
&&&&\\
0716+714		&07 21 53.44	&+71 20 36.36	  &$\sim$ 0.300	&15.6-11.5	\\
MRK\,421		&11 04 27.31    &+38 12 31.79   &\ \ \ 0.031    &13.3-11.7       \\                
BL\,Lac		&22 02 43.29    &+42 16 39.98   &\ \ \ 0.069    &15.3-12.3      \\
&&&&\\
\hline
\end{tabular}
\end{center}
\end{table*}

Light variability in blazars represents a common and  complex phenomenon 
 \citep[e.g.][]{Lindfors2016,Falomo2014,Sandrinelli2014a}. The increase in 
 multi-wavelength observational data recently has provided the 
 possibility to investigate  long-lasting light variations.
Different physical processes occurring in various regions of the accretion 
disk or the jet can co-exist and cause variability signatures with multiple
timescales, making the light curve description a very hard task
a very hard task.
Superposition of these processes to chaotic variabilities, frequency-structured 
noise \citep[e.g.][]{Press1978,Vaughan2005}, and uneven sampling
can overshadow  persistent signals, and the presence of reliable regularities 
become difficult to find. 

The detection of quasi-periodicities could provide inductive support that some 
specific motions may exist in these objects, such as an orbiting secondary black 
hole in a binary black hole system  \citep[BBHS, e.g.][]{Begelman1980,Lehto1996} 
or a binary  torque-warped circumbinary disk \citep[e.g.][]{Graham2015a}, 
both of which possibly result in periodic tidal-induced disk instabilities or  pulsational accretion.
Disk-connected jet flow could simulate these fluctuations.
Geometrical effects, such as  precession of the jet \citep[e.g.][]{Stirling2003},
or helical structures in jets \citep[e.g.][]{Camenzind1992}  correlated to differential
 Doppler boosted flux \citep{Villata1999} can also be inferred.
Short-lived oscillations may be also ascribed to not real long-lasting periodic
processes, which  arise from  shocks along the jet 
\citep[e.g,][]{Marscher1985,Marscher2014} or internal instabilities.
The mounting of observational evidences  of regular oscillations 
in blazars can be successfully explained
with the existence of the above mentioned mechanisms, 
whose physical properties can be constrained through models.  

The \textit{Fermi} mission  has regularly been monitoring the entire sky in the 20 MeV
 - 300 GeV since July 2008. This has enabled us to construct the light curves of 
 several blazars \citep[e.g.][]{Abdo2010}, which are the dominant sources  
 of the extragalactic $\gamma$-ray sky.  
For the brightest sources, it is possible to search for oscillations (quasi-periodicities) 
up to year-like time scales, owing to the limited total sampling time.
The finding of a connected periodicity in the optical band could
greatly enhance the physical relevance of the periodicity itself.
The optical data are, in general, sparse since they are derived from several observatories 
and, contrary to the $\gamma$-ray ones, are very unevenly distributed. 
On the other hand they may extend for various decades.

Up to now  quasi-periodicities have been 
tested in a number of blazars for spurious detections against the red noise background and have 
provided low-significance, yet possibly related, periods in the two bands.
In particular, we mention     
  PKS 2155-304 \citep[][where T$_{\rm \gamma}$= 640 days and T$_{\rm opt} \sim \frac{1}{2}
 T_{\rm \gamma}$]{Sandrinelli2014b,Sandrinelli2016a};  
PG 1553+11 \citep[][where T$_{\rm \gamma}=  T_{\rm opt}=$ 798 days]{Ackermann2015}, 
 and PKS 0537-441 \citep[][where T$_{\rm \gamma}$= 280 d and T$_{\rm opt} \sim
 \frac{1}{2} T_{\rm \gamma}$]{Sandrinelli2016b}. 
 The cases of OJ 287, 3C279, PKS 1510-089, PKS 2005-489 have also been studied.
 We note that in the optical for OJ 287 there is  rather robust evidence of a 12-year 
 oscillation \citep{Sillanpaa1988,Lehto1996},
while  low-significant  quasi-periodities  
T$_{\rm \gamma}$ $\sim$ $T_{\rm opt}$ $\sim$ 400 d \citep{Sandrinelli2016a}
have recently been confirmed in optical by \cite{Bhatta2016}, who also found a  harmonically 
connected peak at  T$_{\rm opt} \sim$800 days.

Among other claims of optical, radio, or optical-radio regular oscillations  of year timescale 
in blazars, we refer to 
 the quasi-periodic radio-band behavior of 
  BL Lac \citep[$\sim$8 yr,][]{Villata2004,Villata2009} 
and   the occurrence  of major  radio-optical outbursts in 
AO 0235+16 \citep[$\sim$5.7 yr or longer,][]{Raiteri2001, Raiteri2006,Raiteri2008} that are  
compatible with the optical light curves.
A helical jet model that leads to changes in Doppler-boosted flux \citep{Ostorero2004}
and driven by  orbital motion in a BBHS was proposed for both the sources.
Quasi-periodical   outbursts 
were also observed   in
0716+714 \citep{Raiteri2003} in optical and radio.  
Evidence of optical quasi-periodicities has been given for 
GB6  J1058+5628 \citep[6.3 yr, ][]{Nesci2010}.
An  oscillatory pattern    of  $\sim$ 150 days  
was detected by \cite{King2013} 
in the BL Lac object J1359+4011 in the 15 GHz wave-band.
On a  similar timescale, a strong and persistent quasi-periodic signal 
 was also  detected in the flat spectrum radio quasar
 PKS\,1156+295 \citep{Wang2014}.
Recently, searching for periodicities 
in inhomogeneous covered light curves,  \cite{Li2016} found oscillations  of  290- to 300-day duration
 in the  X-ray and $\gamma$-ray light curves of MRK 421
at $\sim$ 2.5$\sigma$ over the red noise 
 and a faint signal 
in radio,  which were ascribed to 
helical motions of the emitting material.
In the X-ray domain, a 420- day quasi-periodicity in
 1ES 2321+419 was reported by \cite{Rani2009}. 

Here we examine three more BL Lac objects in the Northern sky:  0716+714, 
MRK\,421,  and BL\,Lac  (see Table \ref{targets}). 
In Sect. 2 we present their $\gamma$-ray and optical light curves. The search for periodicities, 
and the assessment of their significance is given in Sect. 3, and the results are discussed 
in Sect. 4.

\section{Light curves}

\begin{figure}
\centering
\caption{ \label{g1} $Fermi$ light curve  of 0716+714, MRK\,421, and BL\,Lac  
in the 100 MeV-300 GeV energy range (Aug. 2008  - Jan. 2016).} 
\vspace{0.5cm}
 \includegraphics[trim=0.5cm 1cm 0.cm 9.5cm, clip=true,width=0.5\textwidth]{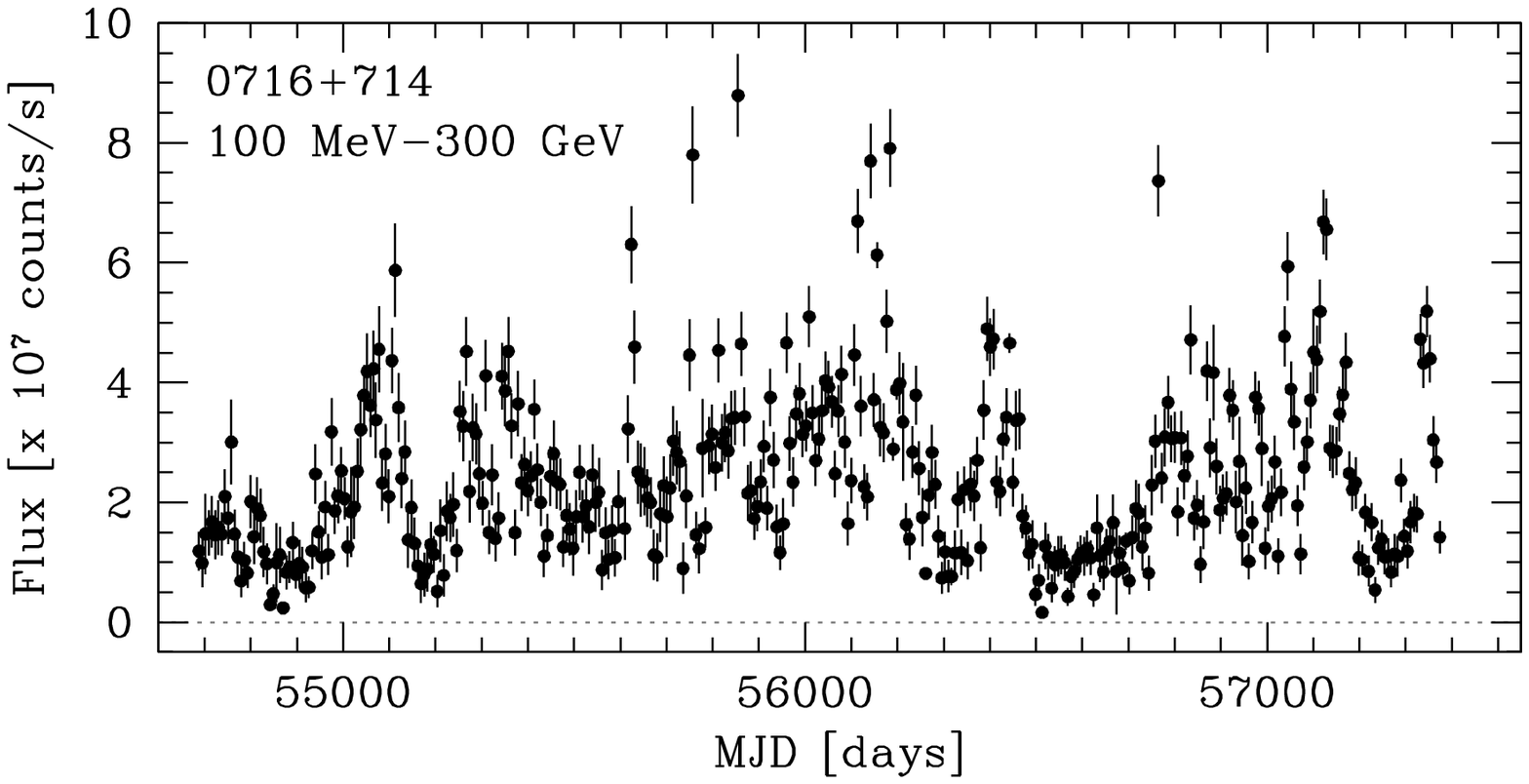} 
\includegraphics[trim=0.5cm 1cm 0.cm 9.5cm, clip=true,width=0.5\textwidth]{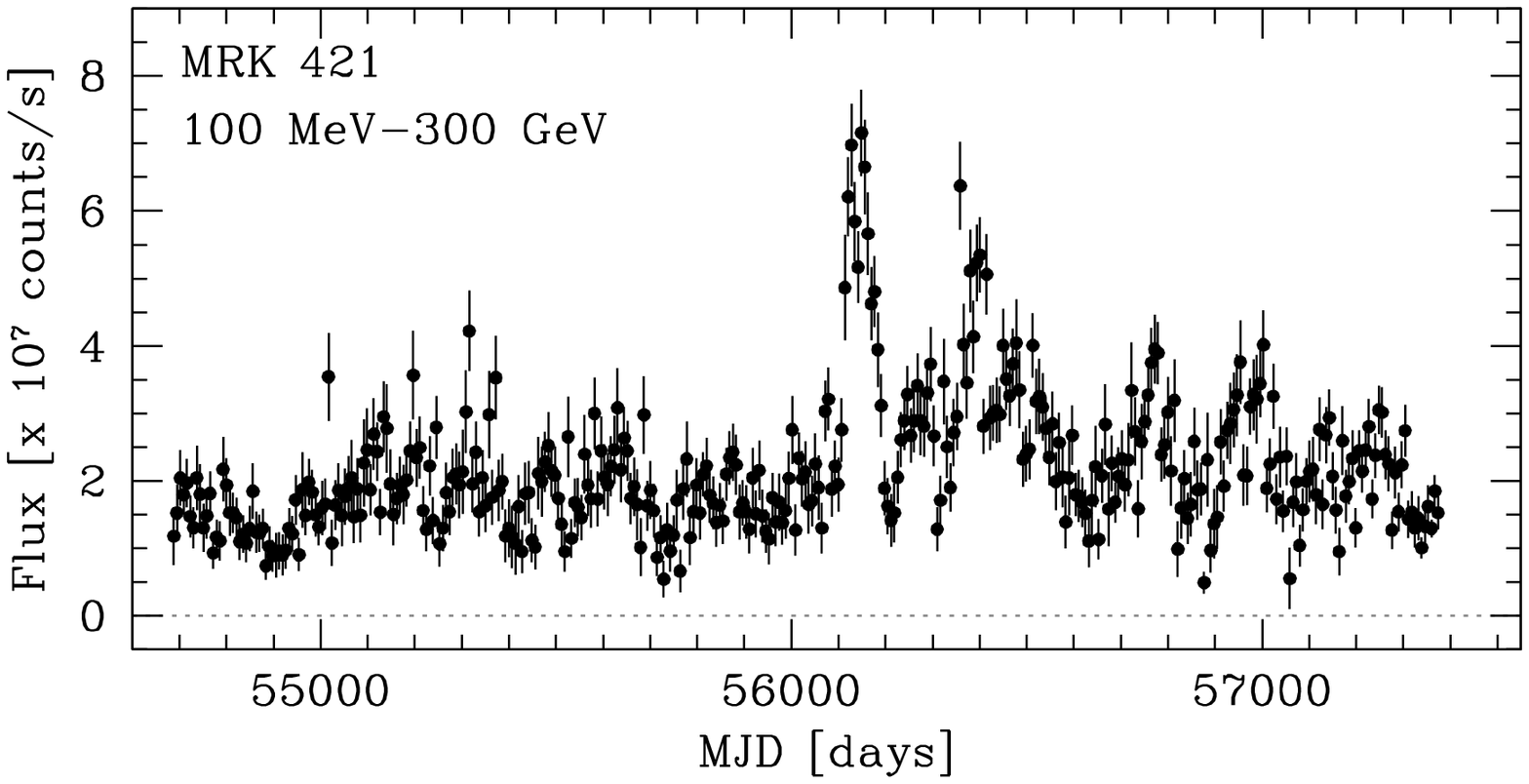} 
\vspace{-1cm}
\includegraphics[trim=0.5cm 1cm 0.cm 9.5cm, clip=true,width=0.5\textwidth]{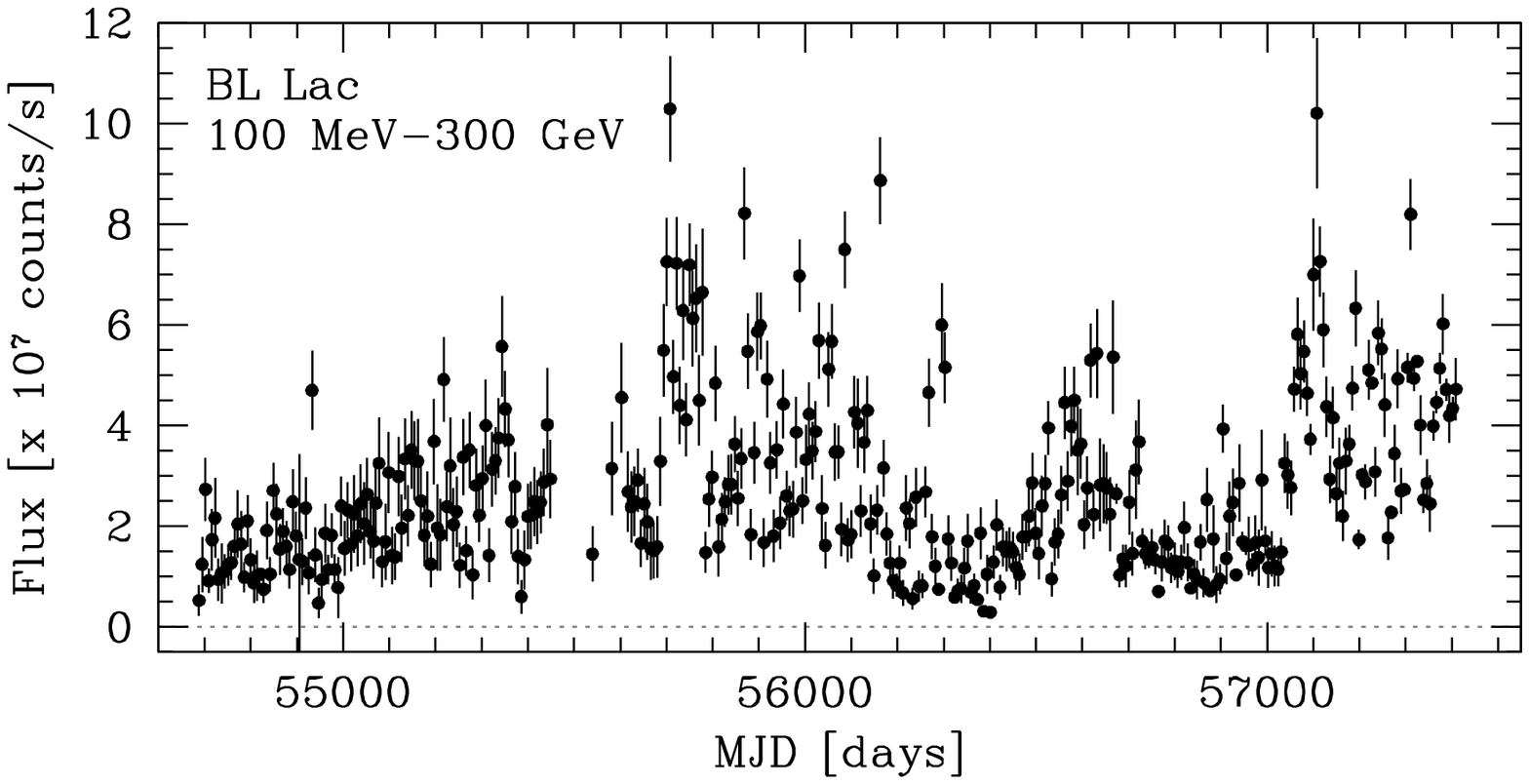} 
\vspace{0.5cm}
\end{figure}

The $\gamma$-ray light curves were taken from the \textit{Fermi}
site\footnote{\texttt{http://fermi.gsfc.nasa.gov/ssc/data/access/lat/msl\_lc/}},
choosing the energy interval 100 MeV - 300 GeV and a binning of 1 week.
 They 
 are reported in Figure \ref{g1}. 
    In all three cases the variability is up to a factor 10,  with presence of flaring 
  episodes superposed on to chaotic, and possibly structured modulation. 
  The fractional variability amplitudes \citep[e.g.][]{Sandrinelli2014a} are
$\sigma^2_{rms} =$ 0.56, 0.33, and 0.59  for 0716+714,  MRK\,421, and BL\,Lac,
respectively.

For the optical bands, we used data 
that was derived from the Tuorla observatory for the three sources.  
The Tuorla Blazar Monitoring Program{\footnote{\texttt{http://users.utu.fi/kani/1m}}} 
 \citep{Takalo2008} was started in 2002 to study the optical behavior of TeV 
candidate blazars reported by \cite{Costamante2002}.
The sample has been extended over the years, 
adding sources that are either detected in very high-energy 
$\gamma$-rays, or are considered as promising candidates. 
A large number of the observations  are 
 performed using the KVA 35cm telescope
at La Palma, which is operated remotely from Finland. 
Additional observations are 
 performed at the 40 cm Searchlight 
Observatory Network telescope, New Mexico, USA and at the 60 cm 
telescope at Belogradchik, Bulgaria. 
The observations are coordinated with the MAGIC Imaging
Air Cherenkov Telescope and, while the monitoring observations
are typically performed two to three times a week,
during MAGIC observations the sources are 
observed every night. The data are analyzed using standard aperture
photometry procedures with the semi-automatic pipeline developed
in Tuorla (Nilsson et al. in preparation).
For 0716+714, MRK\,421, and BL\,Lac, the light curves from Tuorla 
blazar monitoring program extend for more than 13 years but, in this study,
 only data taken during the Fermi mission have been used.
 
\begin{figure}
\centering
\caption{ \label{R}  
Nightly averaged R-band  light curves (Aug. 2008  - Jan. 2016) for 0716+714 and  MRK\,421 
from Tuorla Observatory, and for BL\,Lac 
 obtained combining  nightly averaged  
data from Tuorla  (black) and the WEBT observations (violet). 
The light curves are not host galaxy subtracted.
The reported light curves match with the $Fermi$ monitoring.
 Error bars are, in most cases, smaller than symbol size. 
}
\vspace{0.5cm}
\includegraphics[trim=0.5cm 1cm 0cm 9.5cm, clip=true,width=0.5\textwidth]{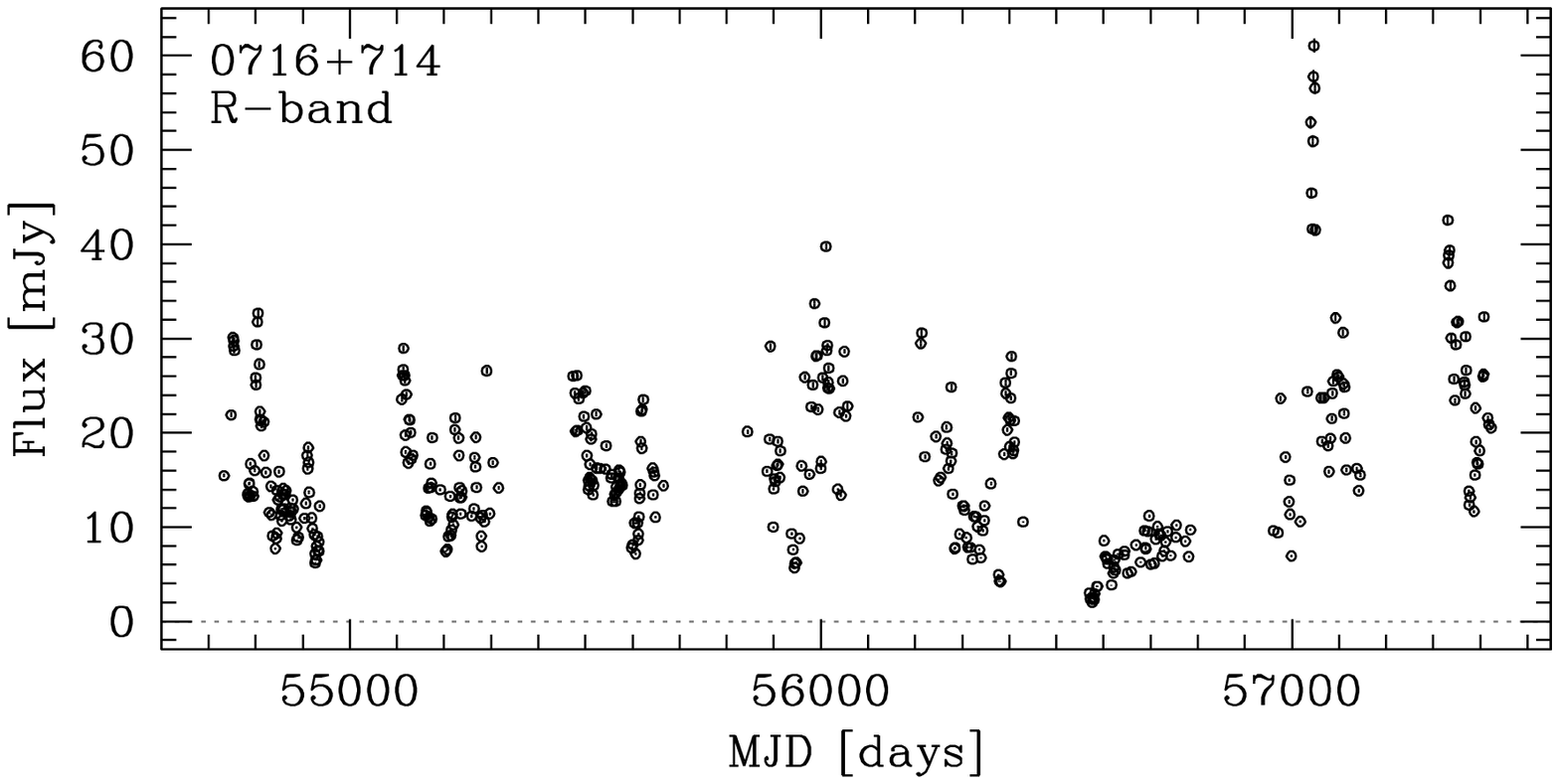} 
\includegraphics[trim=0.5cm 1cm 0cm 9.5cm, clip=true,width=0.5\textwidth]{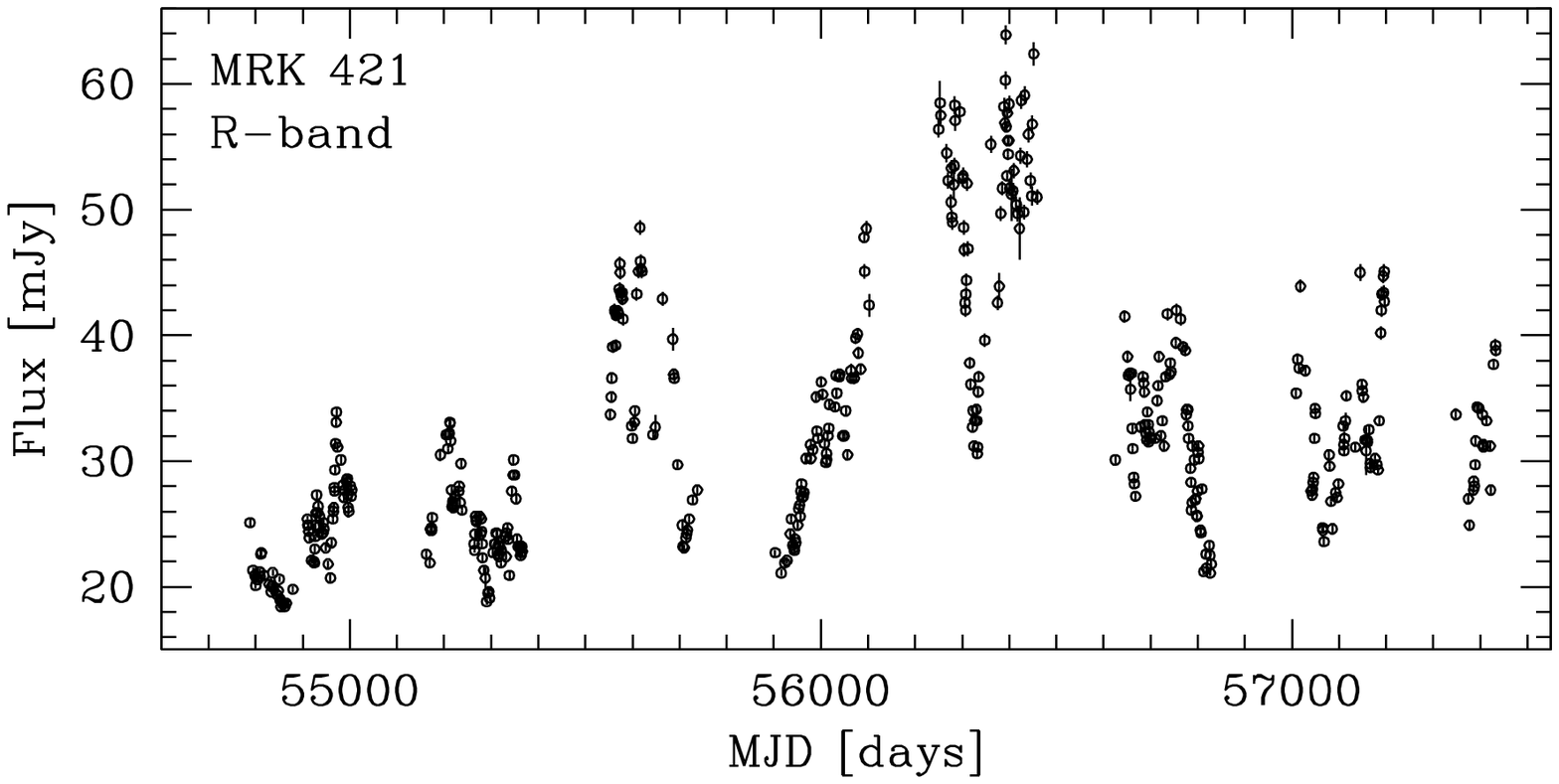} 
\vspace{-1cm}
\includegraphics[trim=0.5cm 1cm 0cm 9.5cm, clip=true,width=0.5\textwidth]{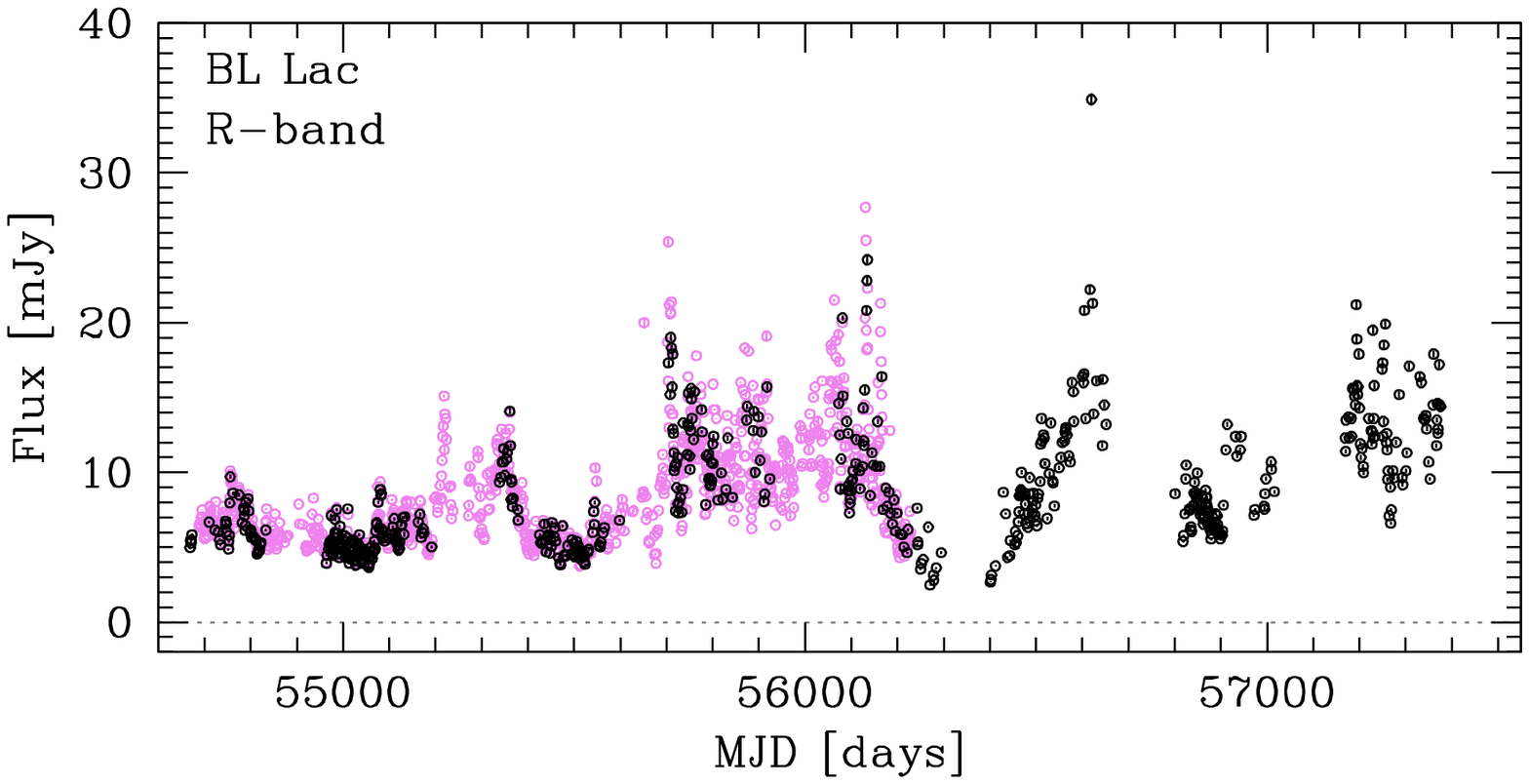} 
\vspace{0.5cm}
\end{figure}

In the case of BL Lac, these data were complemented by those 
that were derived from published material from the Whole Earth Blazar 
Telescope\footnote{\texttt{http://www.oato.inaf.it/blazars/webt}} 
collaboration
 \citep[WEBT, ][]{Villata2002}.
The collaboration  was  established in 1997
 with the aim of organizing observing campaigns to 
investigate blazar variability. Many tens 
of astronomers
taking data in the optical, near-infrared, and radio bands,
are involved in these campaigns, which are usually 
carried out in conjunction with observations at high 
energies by satellites and/or Cherenkov telescopes. 
The data are 
carefully analyzed and assembled to build homogeneous 
light curves. These are finally archived and are available
upon request one year after the publication of the results.
BL Lacertae has been one of the favorite WEBT targets,
since it was followed over several campaigns 
 \citep[see][and references therein]{Raiteri2013}. 
The data used in this paper come from the WEBT archive.

The relevant nightly averaged light curves matching with the $Fermi$ observation span
are reported in Figure\,\ref{R}. 
The variability of MRK 421 and BL Lac is up to 5, and for 0716+714 up to a factor  10.  
The shapes of the light curves are grossly similar to the $\gamma$-ray ones, 
with fractional variability 
$\sigma^2_{rms} =$ 0.53, 0.30, and 0.43  for  0716+714,  MRK\,421, and BL\,Lac, respectively.

\section{Search for quasi-periodicities}

\subsection{Methods}
\label{sec:methods}

Identifying features in periodograms is a delicate task and requires several steps, i.e. 
a careful modeling of the noise and an analysis of the significance with respect to
the noise model.

Essentially, we followed the procedure described in
 \citet{Vaughan2005,Vaughan2010} and \citet{Guidorzi2016}. 
 Power density spectra (PDS) were derived by means of generalized 
 Lomb-Scargle periodograms \citep[e.g.][]{Scargle1982} and, for a 
 cross-check, by means of a completely independent technique 
 \citep{Huijse12,Protopapas15,Huijse15} based on a metric that 
 combines the correntropy (generalized correlation) with a periodic
  kernel to measure similarity among samples separated by a given period. 
  The results obtained with the two techniques are in excellent agreement. 
  We restricted our study to long periods, i.e. in the $50 \le P \le 1000$\,days 
  interval, where the limits are driven by the length of the considered 
  light-curves, and to avoid the very noisy high-frequency part of the PDS.
   All the computations are carried out by procedures coded in  
   {\tt python\footnote{http://www.python.org}} (version 3.5.2) language. 
   Statistical tests and functions, as well as optimization tools, are provided 
   by the {\tt scipy\footnote{http://www.scipy.org}} (V. 0.18.1) library. 
   Lomb-Scargle periodograms are computed by means of the 
   {\tt astropy\footnote{http://www.astropython.org}} (version 1.2.1) 
   implementation. The so-called correntropy periodograms are computed 
   by the {\tt P4J\footnote{https://github.com/phuijse/P4J}} (version 0.13)
    {\tt python} library.  Leahy normalisation was adopted for the 
    analyses of the PDS \citep{Lea83} while, in the figures, we normalized 
    the resulting PDS to be within the range $0 \le P \le 1$ \citep{Zechmeister2009}.

\begin{figure*}[t]
\centering
\caption{ 
\label{singlef_g}   
From left to right, PDS of 0716+714, MRK\,421, 
  and BL\,Lac   from the 100 MeV-300 GeV \textit{Fermi} light curves,
  from  Aug 2008 to Jan 2016.
  Lomb-Scargle spectrum of the input time-series data is given in
   blue and the best fit  noise spectrum in red.  Single frequency  
   95.0\% and 99.7\%  
   false alarm levels are reported by 
   purple and yellow lines.   
   Solid and dashed lines refer to PL and AR1models, respectively.
   }
\vspace{0.3cm}
\includegraphics[trim=0.3cm 0cm 1cm 0cm,clip,width=0.67\columnwidth]{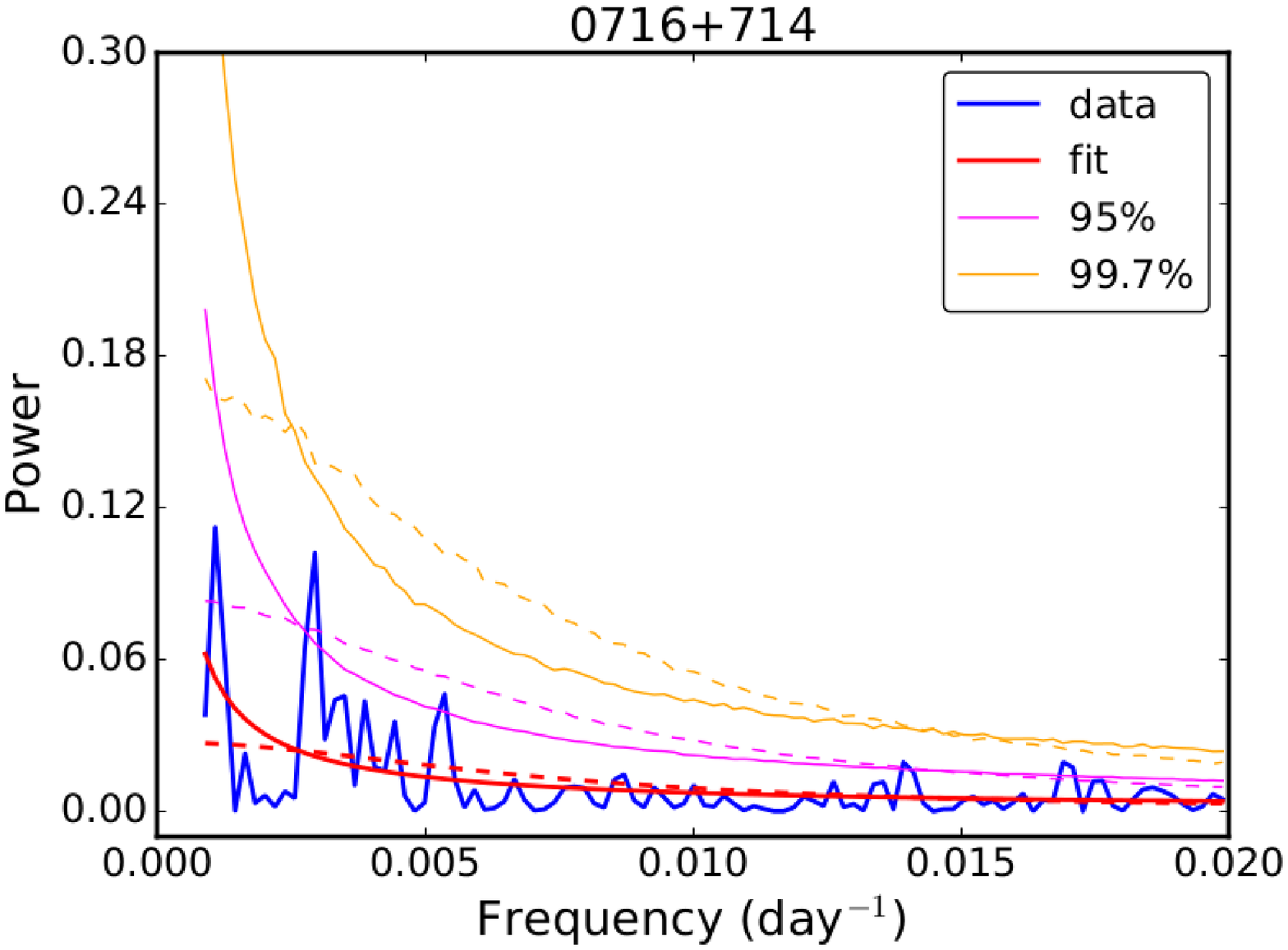} 
\includegraphics[trim=0.3cm 0cm 1cm 0cm,clip,width=0.67\columnwidth]{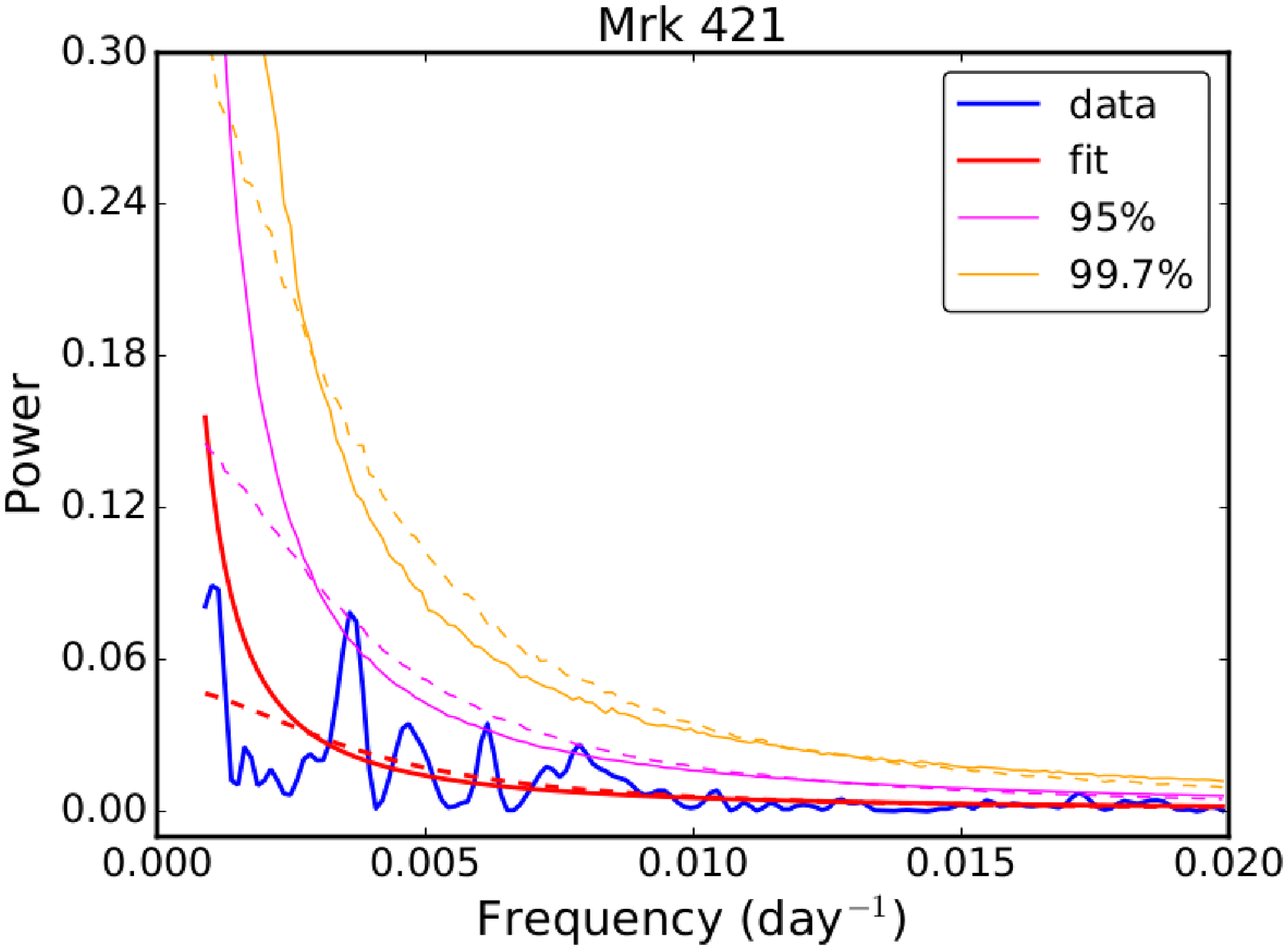} 
\includegraphics[trim=0.3cm 0cm 1cm 0cm,clip,width=0.67\columnwidth]{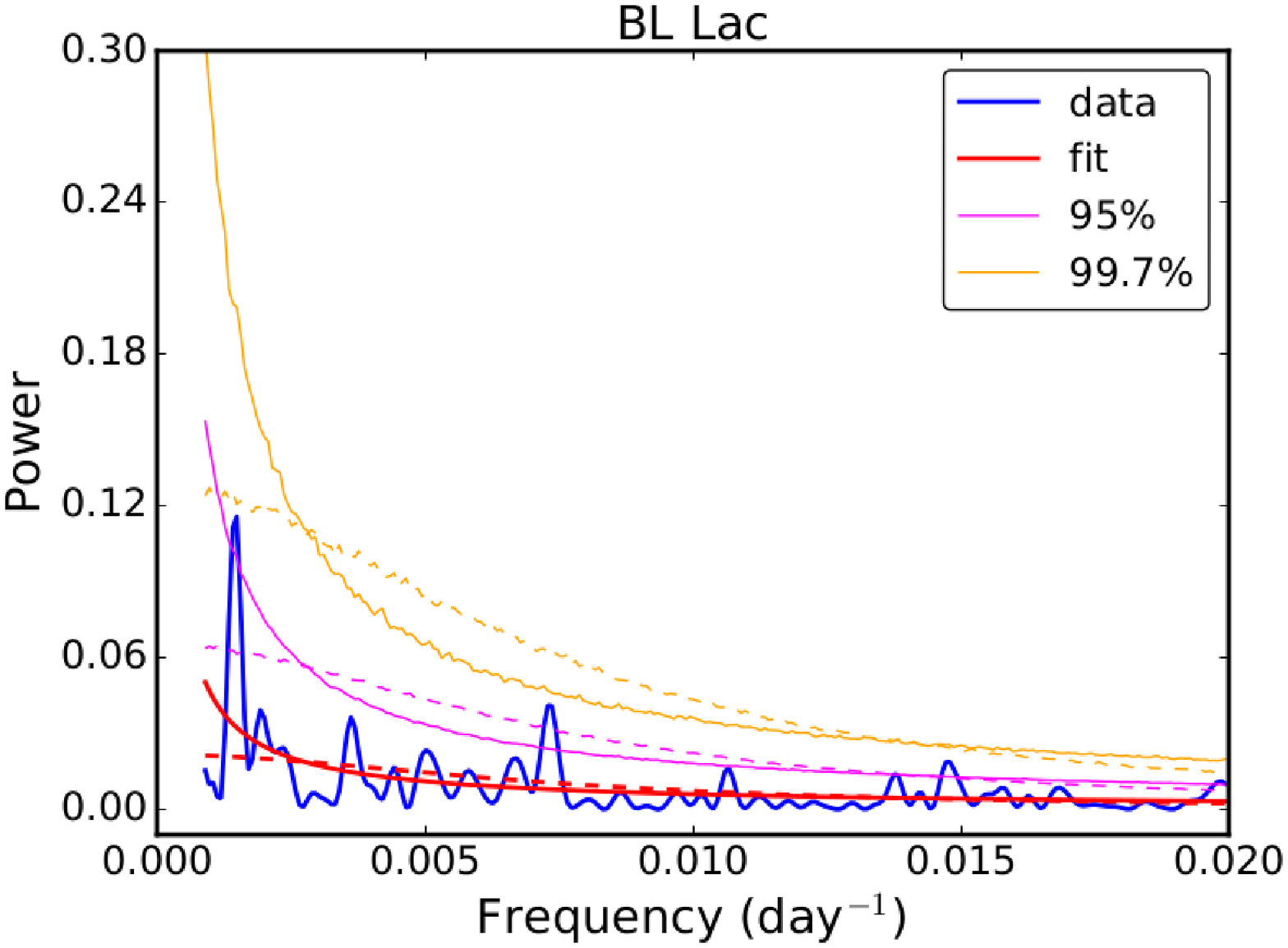}  
\end{figure*}

\begin{figure*}[t]
\centering
\caption{ 
\label{singlef_R}  
  As above  from the R-band  light curves.
  } 
\vspace{0.3cm}
\includegraphics[trim=0.3cm 0cm 1cm 0cm,clip,width=0.67\columnwidth]{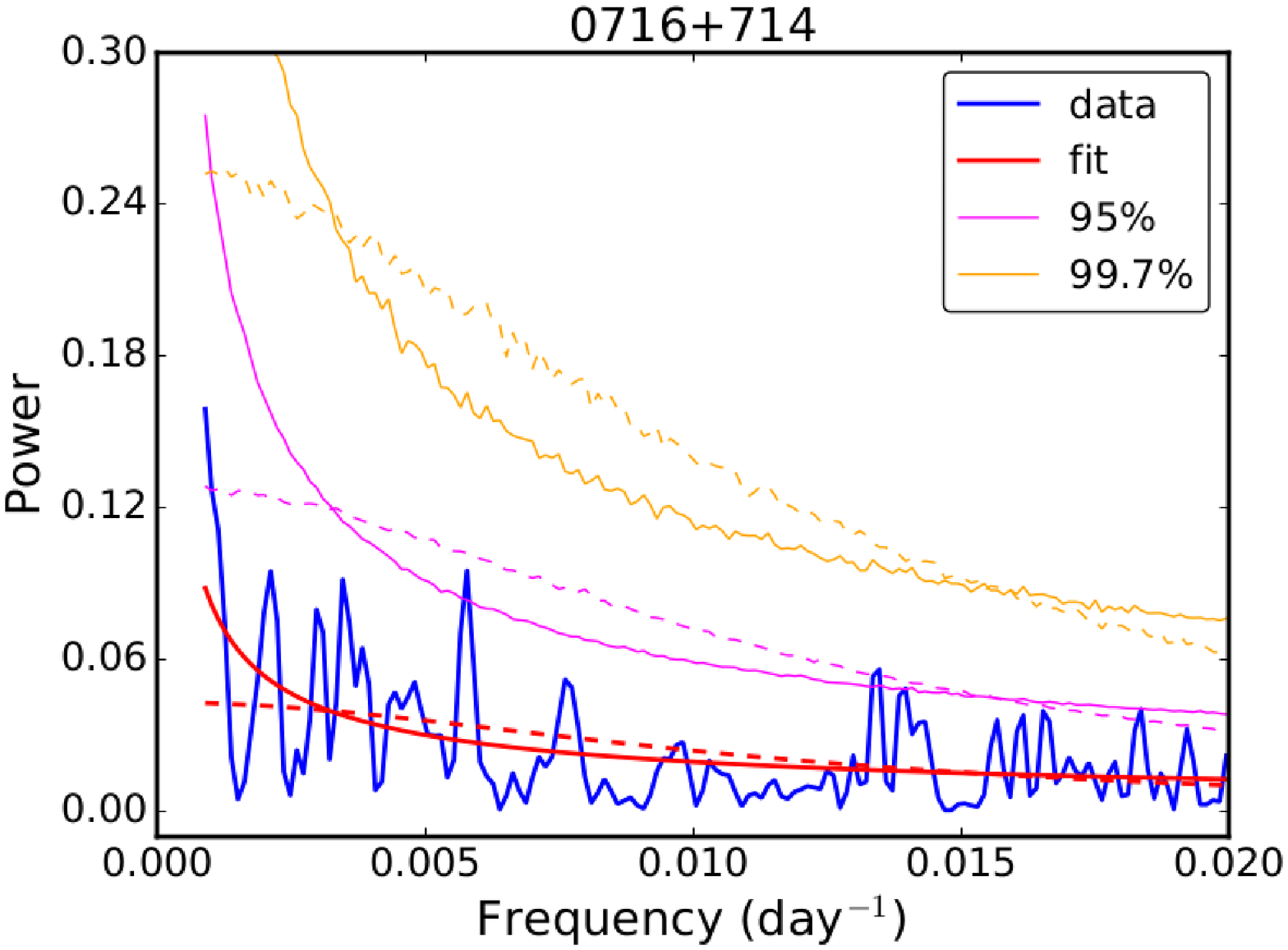} 
\includegraphics[trim=0.3cm 0cm 1cm 0cm,clip,width=0.67\columnwidth]{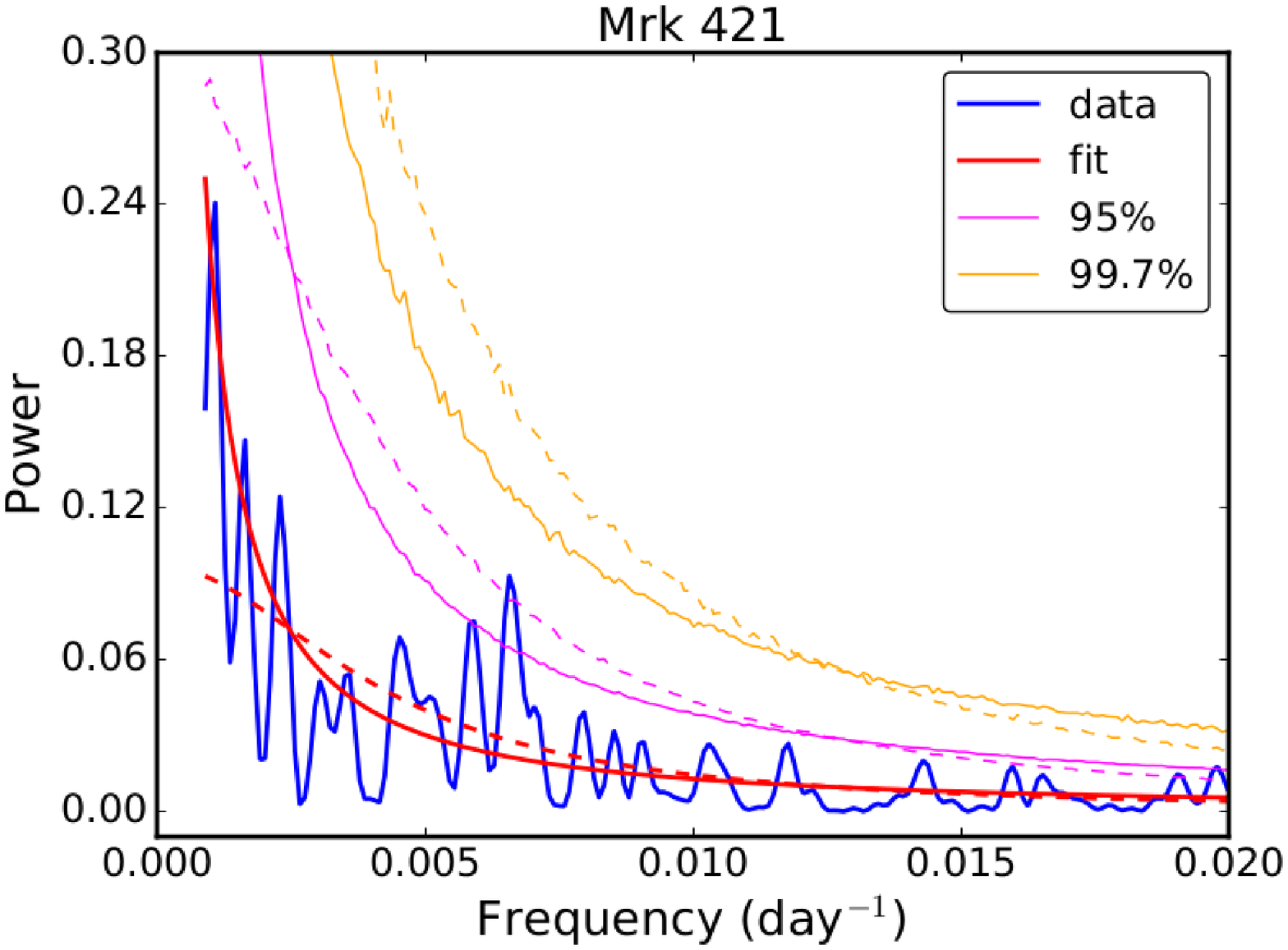} 
\includegraphics[trim=0.3cm 0cm 1cm 0cm,clip,width=0.67\columnwidth]{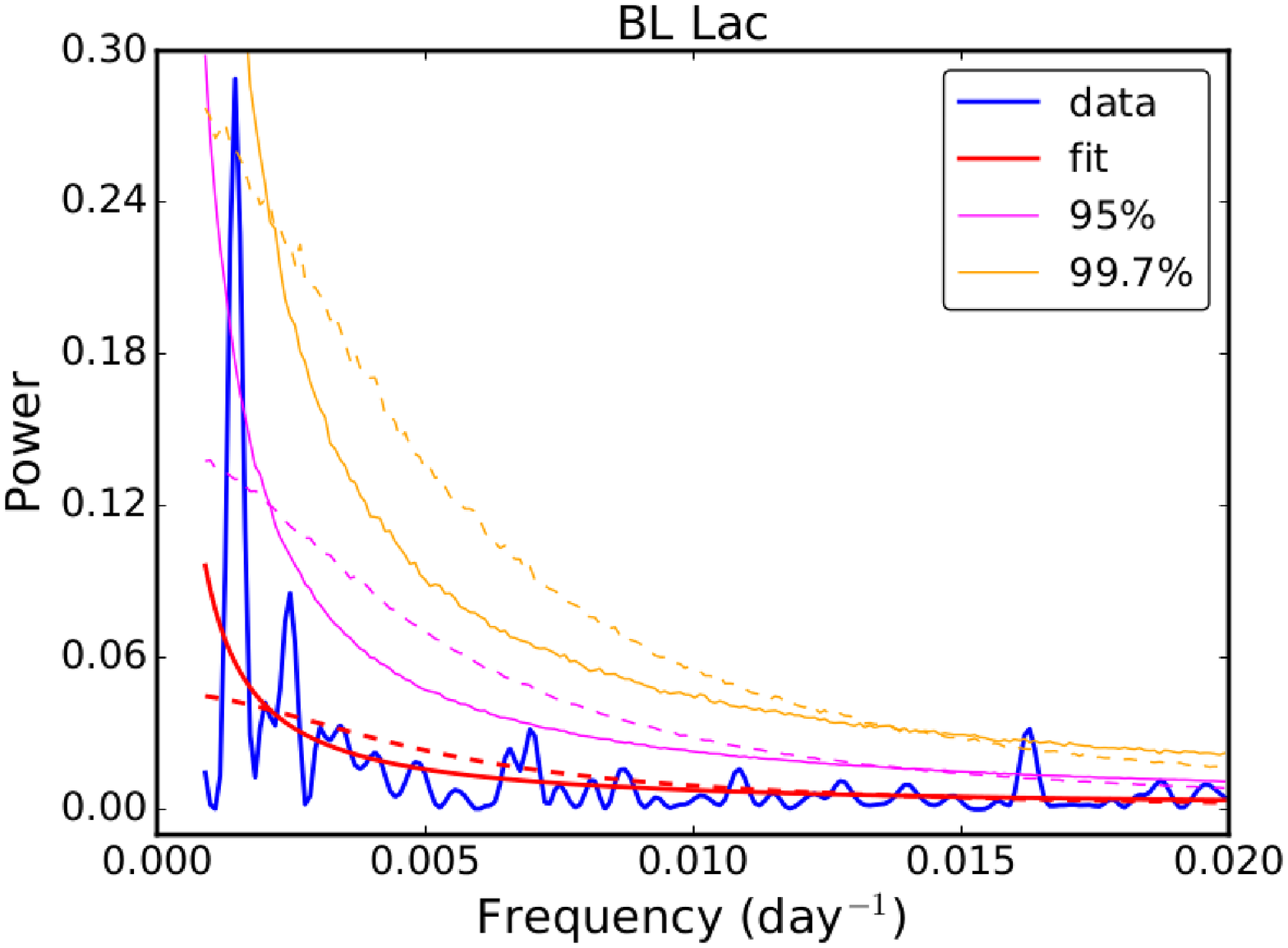}  
\end{figure*}

The noise was modeled as a simple power law (PL) or as an 
auto-regression function of the first order (AR1). Both families of models 
have been considered in the literature \citep[e.g.][]{KoTi97,Koenig1997,
Kelly09,Edelson2013} and, within the limited frequency range of our 
analysis, both provide acceptable fits to the obtained PDS basing on 
Kolmogorov-Smirnov (KS) tests of the residuals against a $\chi^2$ 
distribution \citep{Vaughan2010} with two degrees of freedom. 
A discussion of specific physical meanings for the adopted noise models 
is beyond the scope of this paper yet, as a general rule, PLs tend to predict 
higher noise at the lowest frequencies than AR1 models, although PLs often
 show a more uniform distribution of  residuals. The adopted functional 
 forms  for the PL are
\begin{equation}
S_{\rm PL}(f) = Nf^{-\alpha}, \label{eq:pl}
\end{equation}
where $N$ is the normalization and $\alpha$ the power-law index, always 
positive in our case (i.e. the power is higher at lower frequencies).
For the AR1 model
\begin{equation}
S_{\rm AR1}(f) = \sigma^2 / (1-2a\cos(2\pi f) + a^2), \label{eq:ar1}
\end{equation}
where $\sigma$ is the white-noise variance and $\tau = -1/\ln(a)$ is the 
so-called time constant. $a$ is always positive, again giving more power 
at the lower frequencies. Owing to the limited frequency range 
we studied, the addition of a constant term to the adopted models was never required.

We determine the best–fitting parameters for our PDS in a Bayesian 
framework. We minimized the log-likelihood by means of the so-called
 \textit{Whittle} likelihood \citep{BarVau12}, and the posterior probability density 
 of the parameters of our models are derived by a Bayesian Markov chain 
 Monte Carlo (MCMC) technique. We assumed an uninformative prior 
 distribution (flat or Jeffreys priors).
 We adopted the MCMC 
  implementation provided by the {\tt emcee\footnote{http://dan.iel.fm/emcee/current/}} 
  (version 2.2.1) {\tt python} package, which uses an affine-invariant Hamiltonian 
MCMC.
By sampling the parameters from the posterior distribution, we
derived simulated PDS and  compute the percentiles 
of the simulated periodograms 
at each given frequency (single frequency significances). 
Finally, we also draw the $T_R = \max_j {R_j}$ statistics to 
evaluate the global significance of any peak (i.e. the probability that, 
at  any frequency, the power is equal or larger than a chosen value ) in the PDS \citep[see][for a deeper 
discussion]{Vaughan2010}, where $R = 2P/S$, $P$ the simulated or observed 
PDS, and $S$ the best-fit PDS model. Given that the same procedure is applied
 to the simulated, as well as to the real data, there is no need to perform a multiple
  trial correction owing to the (typically unknown) number of independent sampled 
  frequencies.

\subsection{Results}

\begin{figure*}
\centering
\caption{ 
\label{powg1} 
\textit{Left panel: } PDS of 
 BL\,Lac   from the 100 MeV-300 GeV  \textit{Fermi} light curve  
  from  Aug 2008 to Jan 2016.
  Lomb-Scargle spectrum of the input time-series data is given in
   blue and the best-fit   noise spectrum in red.  Global  
   95.0\% and 99.7\% false alarm levels 
   are shown with  purple and yellow lines.   Solid and dashed lines 
   indicate    PL and AR1 models, respectively.
\textit{Right panel:}  As above from the R-band  light curve. 
}  
\vspace{0.3cm}
\includegraphics[trim=0.3cm 0cm 1cm 0cm,clip,width=0.67\columnwidth]{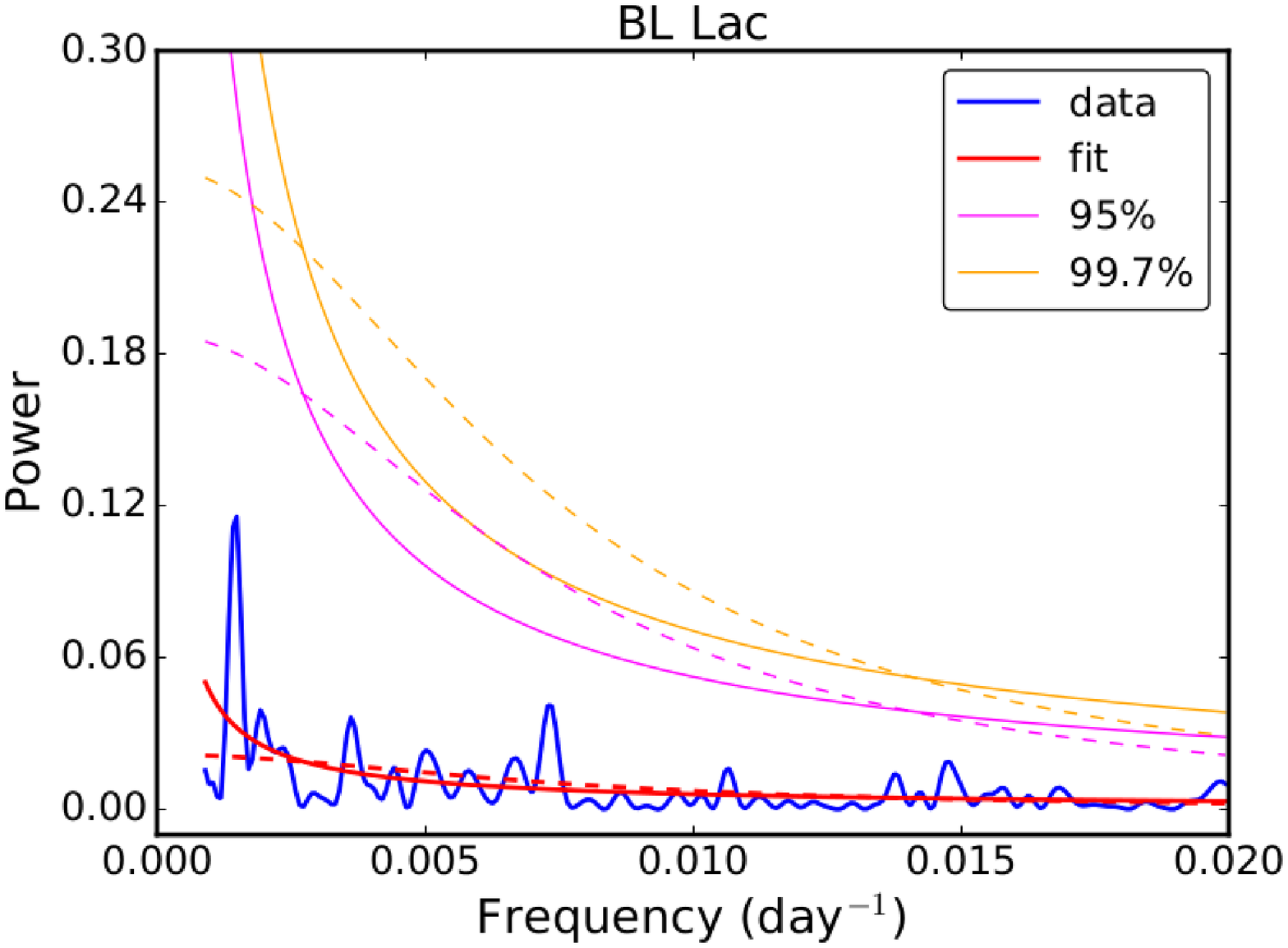}  
\hspace{0.7cm}
\includegraphics[trim=0.3cm 0cm 1cm 0cm,clip,width=0.67\columnwidth]{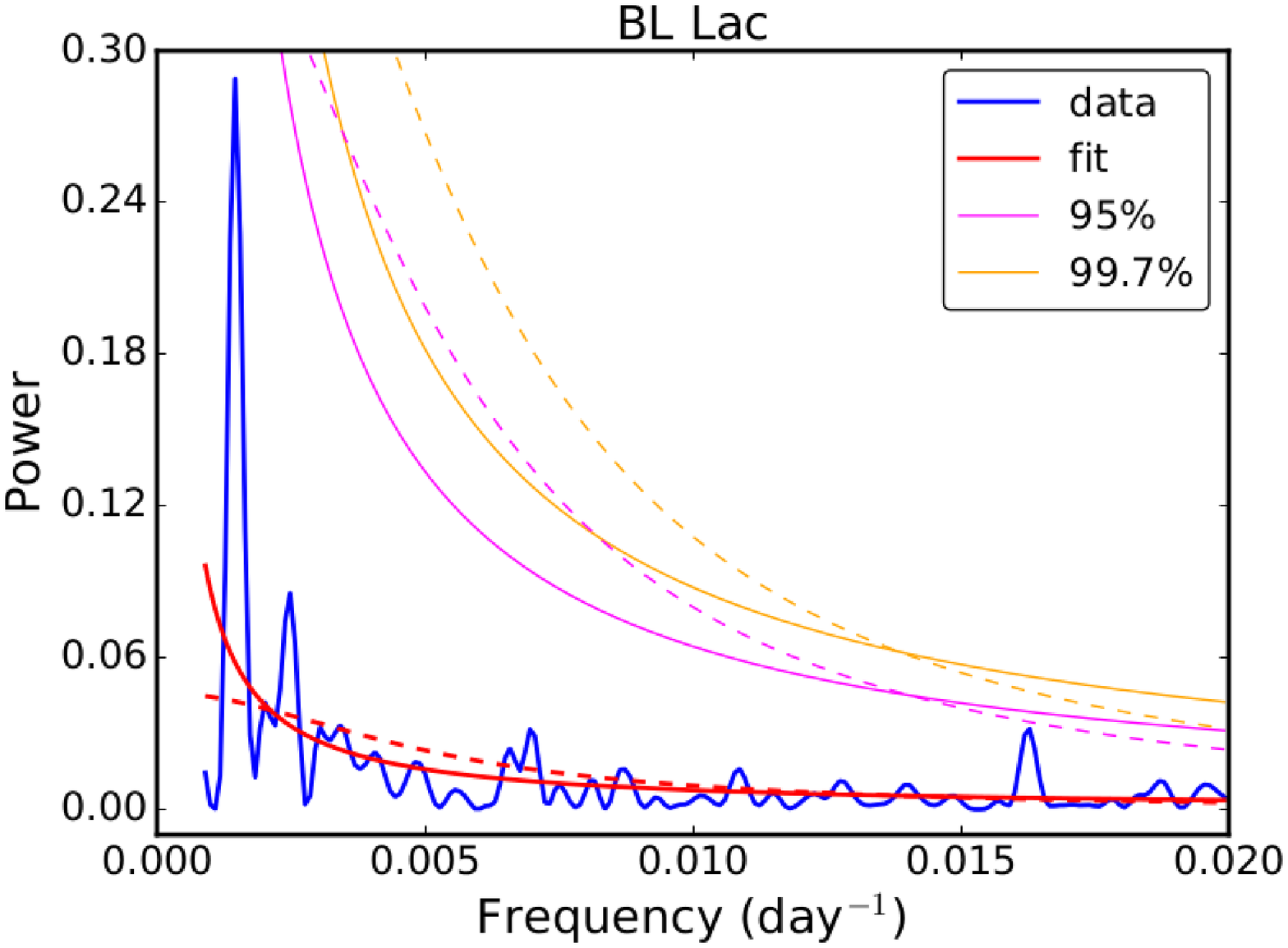}  
\end{figure*}

As a first step, we searched for single frequency Lomb-Scargle peaks 
in the optical and \textit{Fermi} light curves  for the three sources, 
0716+714, MRK\,421, and BL\,Lac,
 using PL and AR1 models, as specified in Subsect. \ref{sec:methods}.
Periodograms given in Figure \ref{singlef_g} refer to the MJD interval 54688-57409 
 (2008 August 10 - 2016 January 22) and correspond to the $Fermi$ light-curves 
 reported in Figures \ref{g1}.
The analysis of  $R$-band  light-curves shown in Figures\,\ref{R},  
binned on 7 days, yielded the periodograms  in Figures \ref{singlef_R}. 
Only in  the case of BL Lac are there 
 peaks with local frequency significances of some interest in both models.
Peaks at 
T$_{\gamma1}\sim$ 680 $\pm$ 35\ days and T$_{opt1} \sim$ 670 $\pm$ 40  days
are identified  at  the same frequency within the errors in the two bands.
The associated uncertainties were calculated 
adapting  the mean noise power level method of \cite{Schwar1991} to red-noise spectra. 
The concurrence  of the same period suggests the  possibility that  the peaks might be related to
a  real quasi-periodicity, although drowned in  very noisy 
 erratic behavior.
A  second peak in the optical  at T$_{opt2}  \sim 60$ days is likely due to the influence 
of the Moon phase on the optical observations. 
We note that, for BL\,Lac in the optical, a peak approximately of the same
 significance also appears  when considering the  sole Tuorla photometry.  
We checked for aliases derived from the interval sampling between the 
 observations or the sampling rate, which can cause false 
peaks in the time analysis \citep[see e.g.][]{Deeming1975}. 
In all the optical curves, there is evidence of a peak at $\sim$ 360 days which is, however,  negligible in the periodograms, with respect to other peaks.

Computing the global (multifrequency) false-alarm probability,
no credible periodicities could be singled out  from the $Fermi$ and optical PDSs  
of the investigated sources. 
The modest significance peak in the BL Lac spectrum at approximately the same
 frequency in the optical and in the $\gamma$-ray band (see Figure \,\ref{powg1}) 
 is still the most interesting feature.

\begin{figure}[t]
\centering
\caption{
Global significances for the peaks in the periodogram  obtained  linearly adding 
 the  $\gamma$-ray and optical PDSs of BL\,Lac.
The Lomb-Scargle spectrum is given in blue. 
Solid and dashed lines indicate PL and AR1 models, respectively. 
Red,  purple and yellow lines are the best-fit noise spectrum,   
95.0\% and 99.7\% false alarm levels, respectively.}
\label{powjoint}
\includegraphics[width=\columnwidth]{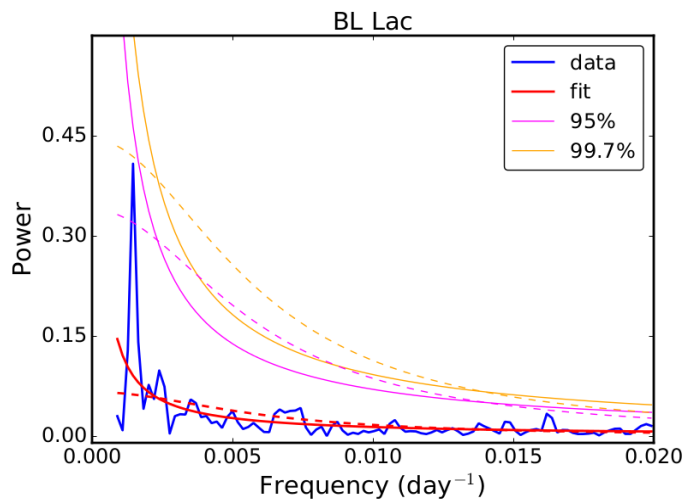}
\end{figure}

To quantify the significance of the detection of the two peaks at the 
same frequency ($T \sim 670-680$\,days) in the optical and $\gamma$-ray bands,
 we apply the same global significance  analysis described in Sect.\,\ref{sec:methods}, 
 adding linearly the PDS obtained for the textbf{$\gamma$-ray} and optical data and 
 evaluating the results against a $\chi^2$ distribution with $2M$ degrees of 
 freedom, where $M$ is the number of added PDS \citep[e.g.][]{BarVau12,Guidorzi2016}. 
 The related PDS peak at $T \sim 680$\,days in Figure \ref{powjoint} now shows  a 
 higher significance, 
near to 99.7 \%,   in the case of the AR1 model while, if we model the noise 
 with a PL, the significance   remains at  90\%.

\begin{figure}
\centering
\caption{ \label{model}  
Comparison between BL Lac light curves in 100 MeV-300 GeV energy range 
and in R-band during the $Fermi$ observational span.
As within the errors $\gamma$-ray and optical peaks are the same,  a
sinusoidal model with the intermediate value of 
T=676\,days is superimposed. 
 Error bars are  omitted for readability.
 }
\vspace{-4cm}
\includegraphics[trim=1cm 3cm 0cm 0cm, clip=true,width=0.5\textwidth]{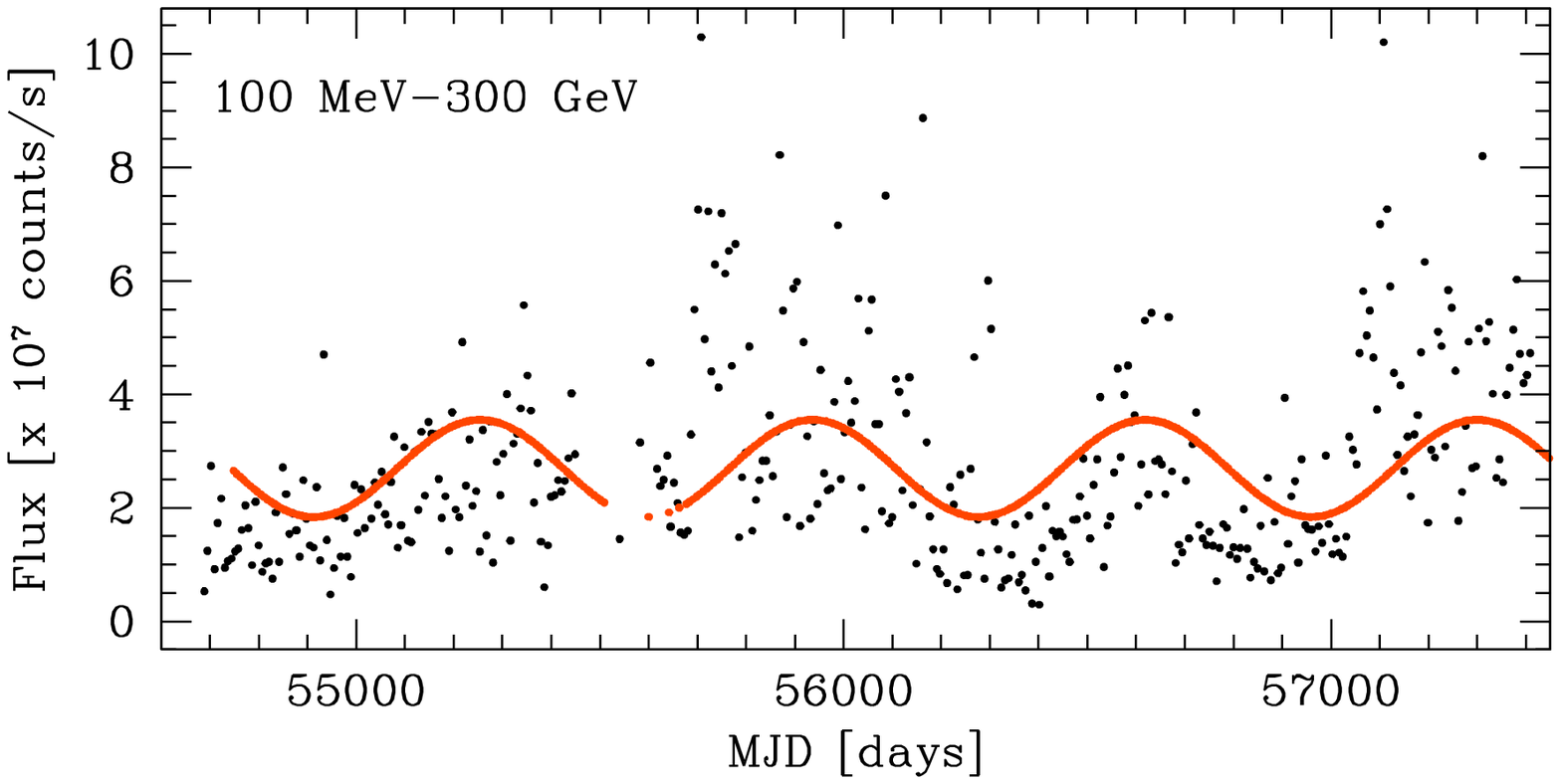} 
\includegraphics[trim=1cm 0cm 0cm 9.75cm, clip=true,width=0.5\textwidth]{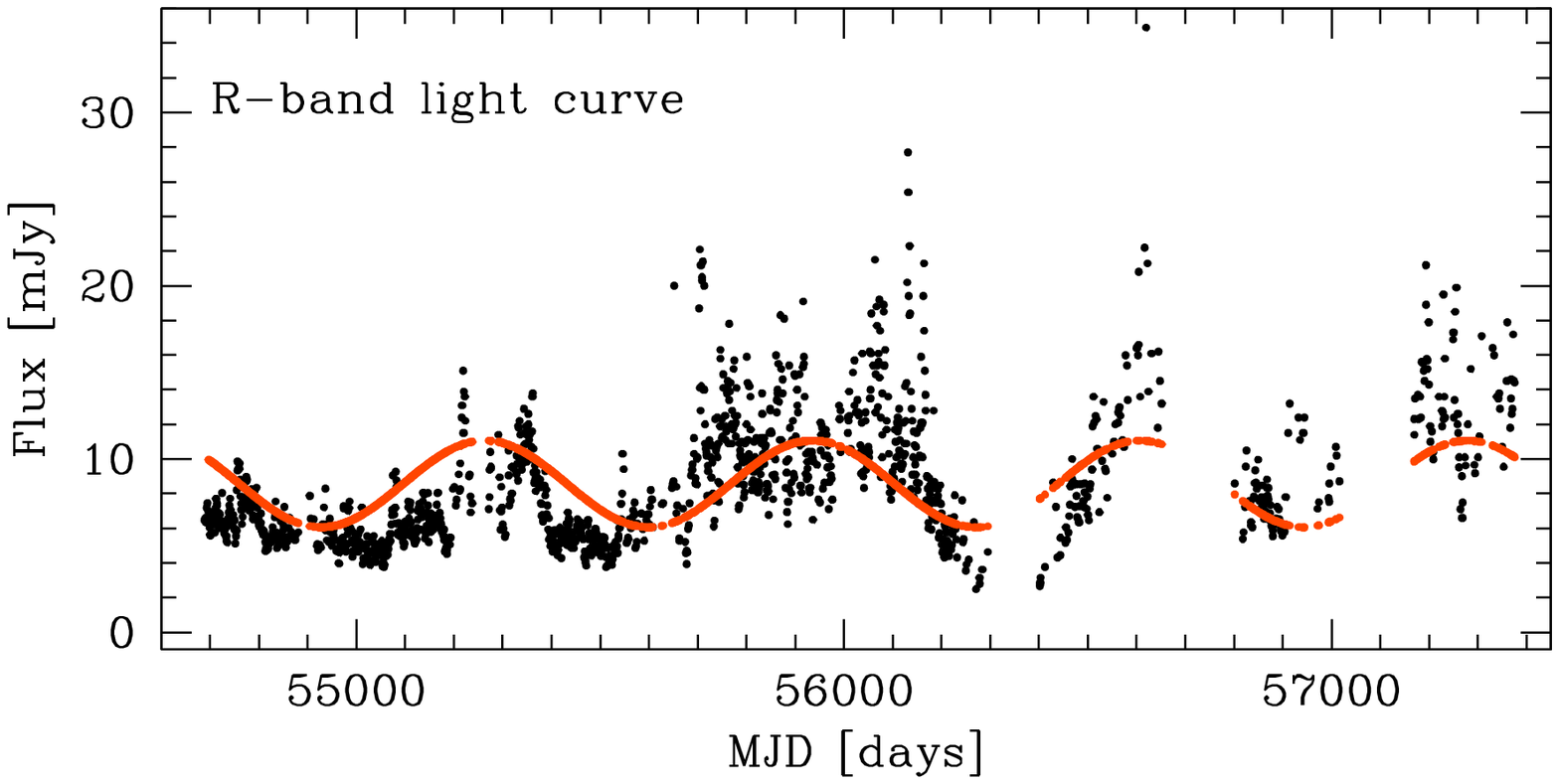} 
\vspace{-1cm}
\end{figure}

\begin{figure}
\centering
\caption{
 \label{cc}  
Discrete cross-correlation function (7-day bin) between 100 MeV-300 
GeV and R-band data for BL Lac.
}
\includegraphics[width=0.9\columnwidth]{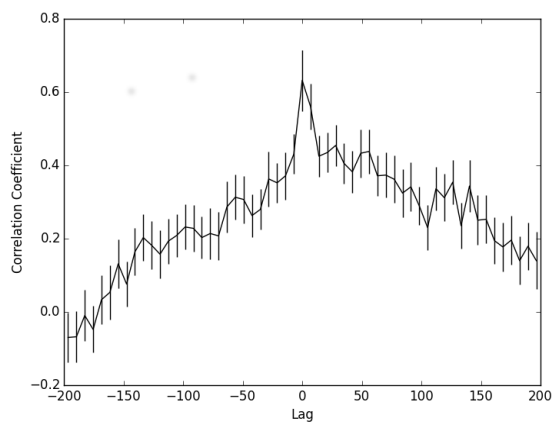}
\end{figure}

In Fig.\,\ref{model}, we finally report  a sinusoidal model for the $\gamma$-ray 
and optical light curves of BL\,Lac, with an intermediate period of 676\,days.  
We computed a cross-correlation of the two light curves, obtained 
following the recipe described by
\citet{Edelson1988} and report it in Fig.\,\ref{cc}.
We note that there is a clear dominant peak at null delay, which indicates a strong 
temporal correlation between the two bands.

\section{ Discussion}

Our analyses show that no highly significant periodicities are present in the 
data we considered.
However, for BL\,Lac, a possible period simultaneously present in the optical 
and in the $\gamma$-rays data sets might be present.  The case of BL Lac therefore appears 
 similar  
to those of PKS\,2155-304, PG\,1553+11, and PKS\,0537-441, to which we have 
referred in Sect.\,\ref{sec:intr}, in the sense that related year-like quasi-periodicities 
are found in both the $\gamma$-ray and optical bands for all these sources. 
As argued in the quoted papers, the oscillations could be directly related to a real periodicity, 
as the orbital one of a system of two binary supermassive black holes, or indirectly as 
being caused by an accretion disc/jet precession.  

The point that we want to emphasize here is that connected year-like 
$\gamma$-ray/optical oscillations might be rather common in bright
\textit{Fermi} blazars. In fact, we considered the light curves of the 72 blazars 
monitored by the \textit{Fermi} satellite from the beginning of its observations, 
or nearly, and reported on the public website\footnote{\texttt{http://fermi.gsfc.nasa.gov/ssc/data/access/lat/msl\_lc/}}. 
Among them, we selected 19 sources that have  a highly significant test statistics 
\citep[TS$>$25,][]{Mattox1996} of the weekly-binned flux for at least  75\% 
of the light curve extension. 
For the quasi-periodicity search, in these light curves we considered  
$\gamma$-rays bins with TS $>$ 4, corresponding to a $\sim$ 2$\sigma$ detection. 
Southern  sources that have the support of our  long-covering  Rapid Eye 
Mounting Telescope 
\citep[REM\footnote{REM data can be retrieved from 
\texttt{http://www.rem.inaf.it}},][]{Zerbi2004, Covino2004} photometry \citep{Sandrinelli2014a}, 
and complemented  with data  drawn from the Small \& Moderate Aperture
 Research Telescope System archives
  \citep[SMARTS\footnote{\texttt{http://www.astro.yale.edu/smarts/glast/home.php}},][]{Bonning2012}, 
  were primarily investigated  \citep{Sandrinelli2014b,Sandrinelli2016a,Sandrinelli2016b}. 
So far  ten systems have been examined in some detail: PKS\,2155-304, 
PG\,1553+11, PKS\,0537-441, BL Lac, OJ287, 3C\,279, PKS\,1510-089, PKS\,2005, 
MRK\,421, and  0716+714 and, in the first four of them, there is some possible
evidence of  related $\gamma$-ray and optical  oscillations. 
Out of the four sources, in three cases (PKS\,2155-304,  PG\,1553+11, 
and BL\,Lac)  we are dealing with BL\,Lac objects, while PKS\,0537+441
may be rather a flat spectrum radio quasar. 

Recently year-like periodicities have been searched for quasars starting
 from the Catalina, Pan-STARRS  and Palomar Transient Factory archives. 
 Examining the light curves of   hundred thousands objects,
only three  apparently well established  periodic cases  have been found
 \citep{Graham2015a,Graham2015b,Zheng2016,Liu2015,Charisi2016},
and $\sim$ 100 are considered as promising candidates. 
As \citet{Vaughan2016} point out, when searching for rare events 
in large samples of noisy time series
it is particularly import  to properly evaluate the false positive rate 
and to treat with caution few-cycle periodicities that are selected by machine
methods. The   tests used in the above   papers
to asses the significance of the claimed quasi-periodicities, which
also lack   calibration  against a red  noise null hypothesis,
could introduce phantom periodic detections.

In any case, there is a huge difference with respect to  bright \textit{Fermi} blazars.  
 The abundance of oscillations in blazars should therefore be related  with the 
 main characteristic that 
 distinguishes them from quasars, the presence of a relativistic jet  pointing 
 in the observer direction.  This would obviously amplify the visibility of the 
 oscillations, since it can magnify the variability through well understood
 relativistic effects and, more importantly, it is responsible of the $\gamma$-ray 
 emission, which, as we have shown, is a main clue for establishing
 the presence of significant oscillations. 
 Furthermore, as already mentioned, it is possible that the oscillations
 are due to jet instabilities, which are independent of the presence of a binary companion. 
 This issue needs to be further explored, especially to clarify if the 
  instability would yield year-like timescales.

If the periodicity were related to a binary system, we would expect a long 
term stability. 
Aiming to cover, say, ten periods, the timescale required to prove  or disprove
the stability of the periodicity would be about a decade.
Therefore, long-term monitoring observations 
 are of great importance to these sources.

\begin{acknowledgements}
We thank the referee for her/his helpful comments on this paper. 
We also acknowledge the critical remarks of an anonymous referee to a previous version
 of the manuscript.
SC  thanks Cristiano Guidorzi for numerous truly enlightening discussions.

 This work has been supported by ASI grant I/004/11/2 and is partly based on data 
taken and assembled by the WEBT collaboration and stored in the WEBT archive 
at the Osservatorio Astrofisico di Torino - INAF\footnote{\texttt{http://www.oato.inaf.it/blazars/webt/}}.
 
This research was also partially supported by Scientific Research Fund from the 
Bulgarian Ministry of Education and Sciences under grant DO 02-137 (BIn-13/09). 
\end{acknowledgements}

\end{document}